\documentclass[aip,floatfix,reprint,twocolumn,superscriptaddress]{revtex4-2}
\usepackage{xcolor} 
\usepackage{graphicx}
\usepackage[hidelinks]{hyperref}
\usepackage{amssymb}
\usepackage[utf8]{inputenc}
\usepackage{amsmath}
\usepackage{ulem}
\usepackage[version=4]{mhchem} 

\DeclareSymbolFont{rsfs}{U}{rsfs}{m}{n}
\DeclareSymbolFontAlphabet{\mathscrsfs}{rsfs}

\usepackage{braket}

\renewcommand{\vec}[1]{\boldsymbol{#1}}

\definecolor{vastkust}{RGB}{0, 48, 80} 
\hypersetup{
  colorlinks,
  citecolor=vastkust,
  linkcolor=vastkust,
  urlcolor=vastkust}

\begin{document}

\title{Fluctuation-induced quadrupole order in magneto-electric materials}

\author{Finja\ Tietjen}
\affiliation{Department of Physics, Chalmers University of Technology, 412 96 G\"{o}teborg, Sweden}

\author{R. Matthias Geilhufe}
\affiliation{Department of Physics, Chalmers University of Technology, 412 96 G\"{o}teborg, Sweden}

\date{\today}
\begin{abstract}
Phases that go beyond dipolar ordering and into multipolar ordering have recently been observed in magneto-electric materials. The resulting phase diagram is commonly explained using the concept of
competing orders and exact microscopic interactions. 
In contrast, we propose an approach based on composite order emerging from a parent phase to explain quadrupoling above magnetic or electric dipolar orders. 
We include thermal fluctuations and symmetry and show their influence on the emergence of quadrupolar order. 
We find an analytical expression for the quadrupolar transition temperature, the critical anisotropy and explain the coupling of the quadrupolar order to mechanical strain, in agreement with experiments. The shift in perspective on quadrupolar ordering from competing to composite order is universal and can be extended to other types of multipolar ordering. 
This offers the possibility of understanding tunability and material-specific predictions of the related phase transitions without explicit knowledge of the microscopic mechanisms.
\end{abstract}
\maketitle

\section{Introduction}

A phase defines the state of a system and determines its functional properties.
Most known and well-studied phases of matter arise from emergent phenomena involving simple constituents, such as charge or spin dipoles or electron pairs.
Recently, however, a growing number of more complex quadrupole and multipolar phases have been observed in both conventional and superstates of matter~\cite{svistunov2015superfluid,fernandes2019intertwined,balatsky1993even,babaev2004phase}.

For instance, in magnetoelectric materials, quadrupole orders have been successfully described using the multipole expansion~\cite{spaldin2008toroidal,ederer2007towards,spaldin2013monopole,thole2018magnetoelectric}.
Rather than treating these complex orders as individual competing phases within a single material, we propose that they may emerge as composite or intertwined orders, similar to the intricate phase diagram of high-$T_c$ superconductors~\cite{aeppli2020hidden,fradkin2015colloquium,fernandes2019intertwined}. 
In fact, the magnetoelectric tensor, with components $\langle r_i \mu_j \rangle$, exemplifies such a composite order derived from a multipole expansion~\cite{thole2018magnetoelectric}.
Notably, its components can be nonzero even in the absence of dipoles $\langle \vec{r} \rangle$ and magnetic orders $\langle \vec{\mu} \rangle$, giving rise to terms that describe magnetoelectric monopoles~\cite{spaldin2013monopole}, toroidal moments~\cite{spaldin2008toroidal}, and quadrupoles.

While the first evidence for quadrupoling in superstates has only recently emerged~\cite{cho2020z3,grinenko2021state,zheng2025counterflow}, magnetic quadrupoling has been observed with strong experimental support.
For example, in \ce{CeB6}, a quadrupole phase transition is detected via X-ray scattering experiments at $T_q \sim 3.2$~K, followed by an antiferromagnetic state below $T_c \sim 2.3$~K~\cite{nakao2001antiferro}.
In \ce{Ba2NaOsO6}, a double perovskite and magnetic Mott insulator, a quadrupole phase transition is observed using nuclear magnetic resonance spectroscopy at $T_q \sim 10$~K, which precedes a ferromagnetic phase transition at $T_c \sim 6.3$~K~\cite{lu2017magnetism}.
X-ray diffraction and scattering experiments on \ce{Ba2MgReO6} reveal that the quadrupole phase, entering at $T_q \sim 33$~K, is linked to a lattice distortion that persists into the magnetic phase below $T_c \sim 18$~K~\cite{hirai2020detection,Soh2024}.
Similarly, heat capacity measurements and X-ray diffraction on single-crystalline \ce{Cs2TaCl6} show two phase transitions at $T_q \sim 15$~K and $T_c \sim 4.6$~K, accompanied by a structural change from cubic to tetragonal at $T_q$~\cite{mansouri2023charge}.

Current theoretical approaches rely on microscopic theories, including \textit{ab initio} calculations~\cite{suzuki2018first}.
For \ce{CeB6}, the multipoles are explained using a microscopic model of electron shells, and experimental measurements are reproduced numerically~\cite{santini2009multipolar}.
In double perovskites such as \ce{Ba2NaOsO6}, spin-orbit entangled electrons account for the observed symmetry in the quadrupolar phase, particularly for systems governed by $5d$ electrons~\cite{hirai2020detection}.
For \ce{Ba2MgReO6}, an \textit{ab initio} approach reveals that the quadrupole phase arises from a combination of electronic and lattice coupling mechanisms~\cite{fiore2024interplay}.
This approach also highlights a strong dependence on uniaxial strain, consistent with experimental observations~\cite{hirai2020detection,Soh2024}, although the predicted transition temperatures are overestimated by $40\%$~\cite{fiore2024interplay}. 
While existing theories often assume competing orders, we demonstrate how quadrupoling can emerge as a composite order.
To this end, we employ a mesoscopic approach based on a coarse-grained free energy.

While our approach is general, we demonstrate its application using the example of cubic symmetry. 
We consider a vectorial order parameter $\vec{X}$, representing, for example, magnetization or polarization. 
This phase is referred to as the "parent phase". 
Our approach captures several qualitative features consistent with recent experiments:
\begin{itemize}
    \item Quadrupole order emerges near a dipole order.
    \item The quadrupole phase transition temperature $T_q$ exceeds the dipole transition temperature $T_d$, but is bounded by $T_d < T_q < 2T_d$ in cubic systems.
    \item The quadrupole order couples linearly to elastic stress and strain, triggering a structural phase transition.
    \item The emergence of quadrupole order is closely tied to the strength and nature of magnetic or electric anisotropy in the material.
\end{itemize}

The free energy density of the parent phase in a cubic system, $f(\vec{X}, T)$, is detailed in Sec.~\ref{sec:free-energy}.
Above the critical temperature $T_c$ of the parent phase, the order parameter is zero, i.e., $\langle\vec{X}\rangle = 0$. However, we show that a composite order described by a quadratic order parameter $\Phi = \langle X_i X_j \rangle$ can be nonzero. 

In Sec.~\ref{sec:phenomenological}, we discuss the dynamics of a multicomponent order parameter $\langle\vec{X}\rangle$ and show that thermal fluctuations increase as $\langle X_i^2 \rangle \sim k_B T$. 
When the system's anisotropy exceeds a critical value, we motivate that non-trivial minima in the free energy landscape might favor a specific type of composite order.  

To derive a free energy of the composite order parameter, we transform the free energy of the parent phase into the free energy of the quadratic order parameter, $F[\vec{X}] \rightarrow F[\vec{\Phi}]$, using the partition function and Hubbard-Stratonovich transformation (Sec.~\ref{sec:quadrupoling}).
By solving the self-consistent equations $\delta F[\vec{\Phi}] / \delta \Phi = 0$ in Sec.~\ref{sec:gap-equation}, we derive a self-consistent equation for the composite order parameter. 
This allows us to analytically determine the transition temperature associated with the quadrupole phase.

To further analyze the behavior of the quadrupole phase, we explicitly derive its free energy up to fourth order in Sec.~\ref{sec:quadru-free-energy}. 
Due to the emergence of a third-order contribution, we conclude that the phase transition would be of first order. 
Also, the symmetry of the quadrupole order parameter suggests a coupling to shear strain and tetragonal distortion in the material, which we explain phenomenologically in Sec.~\ref{sec:strain}.
We validate our approach by predicting the experimental results for the lattice distortion in \ce{Ba2ReMgO6} in Sec.~\ref{sec:BaReMgO}.

\section{Free energy of the parent phase}
\label{sec:free-energy}
According to the Landau-Ginzburg-Wilson theory, a phase transition is characterized by a macroscopic order parameter that captures the essential physical property of the ordered state, breaking the symmetry of the unordered normal state.
We develop our theory of quadrupoling using the example of a vectorial order parameter, also known as an $O(n)$ theory.
In ferro- and ferrimagnets, the magnetization $\vec{M}$ serves as the vectorial order parameter.
For antiferromagnets, where the magnetization is staggered, the Néel vector $\vec{L}$—the difference between domains of opposite magnetization—acts as the order parameter.
In more complex magnetic structures, such as helical magnets, the spiral axis orientation can be described by a vector.
Similarly, local electric dipoles can give rise to ferroelectricity, with the polarization $\vec{P}$ as the order parameter.
An analogous construction to that of antiferromagnets can also be applied to describe antiferroelectric order~\cite{singh1995structure,hao2014comprehensive,toledano2016theory}.

In what follows, unless specified otherwise, we use $\vec{X}$ to denote the vectorial order parameter, which may represent any of the examples mentioned above.
While we argue that quadrupoling can emerge in various crystalline symmetries, we develop our theory using the example of a crystal with cubic symmetry at high temperatures.
We begin by formulating the free energy density $f(\vec{X}, T)$ in terms of the order parameter $\vec{X}$~\cite{geilhufe2024composite,barabanov2022landau,geilhufe2018gtpack,devonshire1954theory},
\begin{multline}\label{eq:freeEn}
    f(\vec{X}, T) = f_0(T) + \vec{X} \left( -c \nabla^2 + r \right) \vec{X} + \beta_+ \vec{X}^4 \\
    + \frac{\beta_-}{2} \left([\vec{X} \mathbb{M}_1 \vec{X}]^2 + [\vec{X} \mathbb{M}_2 \vec{X}]^2\right)\, .
\end{multline}
Here, $f_0(T)$ represents a temperature-dependent contribution to the free energy density, independent of the order parameter.
Although $f_0(T)$ does not influence the formation of quadrupole order, we include it for completeness.

The gradient term $\sim \vec{X} \Delta \vec{X} = -(\nabla \vec{X})^2$ accounts for fluctuations in the order parameter, treating it as a field.
In cubic symmetry, additional derivative terms are allowed, such as those involving the curl $\sim (\nabla \times \vec{X})^2$ and the divergence $\sim (\nabla \cdot \vec{X})^2$.
However, in the absence of magnetic monopoles ($\nabla \cdot \vec{M} = 0$) and assuming no free charges ($\nabla \cdot \vec{P} = 0$), we generalize $\nabla \cdot \vec{X} = 0$.
Similarly, in the absence of local currents and external time-dependent fields, we set $\nabla \times \vec{X} = 0$.
The conventional second-order term is parameterized as $r = \alpha(T - T_c)$.

Due to cubic symmetry, the fourth-order term in the free energy density splits into two contributions, parametrized by $\beta_+$ and $\beta_-$.
The parameter $\beta_+$ corresponds to the fully isotropic term $\sim \vec{X}^4$, while $\beta_-$ characterizes the cubic anisotropy.
We express this anisotropy using the basis functions of the two-dimensional irreducible representation $E_g$ of the cubic group $O_h$,
\begin{align}
    \vec{X} \mathbb{M}_1 \vec{X} = \Phi_{E_{g;1}} &= \frac{1}{\sqrt{2}} (\vec{X}_x^2 - \vec{X}_y^2), \\
    \vec{X} \mathbb{M}_2 \vec{X} = \Phi_{E_{g;2}} &= \frac{1}{\sqrt{6}} (\vec{X}_x^2 + \vec{X}_y^2 - 2\vec{X}_z^2),
\end{align}
where we introduce the matrices
\begin{equation}
    \mathbb{M}_1 = \frac{1}{\sqrt{2}} \begin{bmatrix}
    1 & 0 & 0 \\
    0 & -1 & 0 \\
    0 & 0 & 0
    \end{bmatrix}, \quad
    \mathbb{M}_2 = \frac{1}{\sqrt{6}} \begin{bmatrix}
    1 & 0 & 0 \\
    0 & 1 & 0 \\
    0 & 0 & -2
    \end{bmatrix}
\end{equation}
for notational convenience.

\section{Phenomenological argument for fluctuation-induced quadrupoling}
\label{sec:phenomenological}
Near equilibrium, the order parameter $\vec{X}$ relaxes to its ground state configuration according to the time-dependent Ginzburg-Landau-Wilson model~\cite{Hohenberg1977}, described by the overdamped dissipative dynamics

\begin{equation}
    \dot{\vec{X}}(\vec{r}) = - \Gamma \frac{\delta f(\vec{X},T)}{\delta \vec{X}} + \sqrt{2 \Gamma k_B T}\, \vec{\xi}(\vec{r},t).
\end{equation}

The noise term $\vec{\xi}(\vec{r},t)$ represents uncorrelated Gaussian white noise, satisfying $\left<\xi\right> = 0$ and the correlation function $\left<\xi_\alpha(\vec{r}',t')\xi_\beta(\vec{r},t)\right> = \delta(\vec{r}-\vec{r}')\delta(t-t')\delta_{\alpha\beta}$.
Although this model is stochastic—producing random realizations of $\vec{X}$ at each time step—it ensures that the distribution of $\vec{X}$ follows the Boltzmann distribution.

Above the ferroic transition temperature ($T > T_c$), the average order parameter vanishes, $\left<\vec{X}\right> = 0$, while its fluctuations grow linearly with temperature, $\left<\vec{X}^2\right> \sim k_B T$.
This behavior mirrors that of a Brownian particle in a trapping potential, where the average position is zero, but the average potential and kinetic energies are each $k_B T/2$ due to the equipartition theorem.

\begin{figure}[t]
    \centering
    \includegraphics[width=\linewidth]{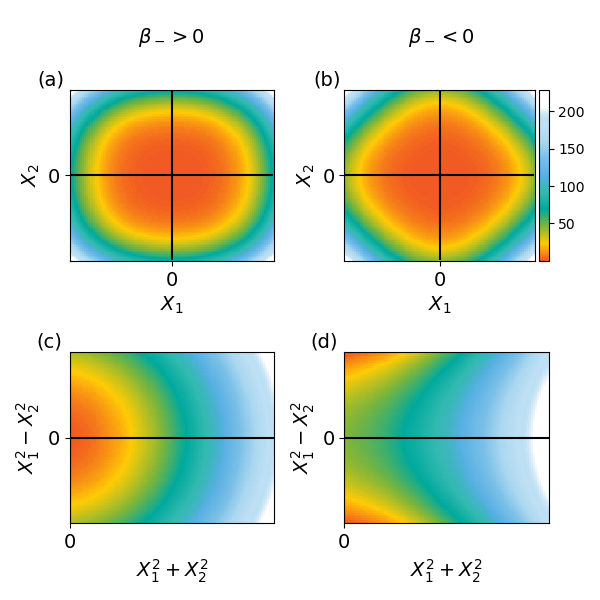}
    \caption{Free energy density minima for linear and quadratic order parameters.
    The quadratic order parameter, representing a composite order (a), exhibits a minimum similar to the linear order parameter (b) at low anisotropy.
    At higher anisotropy, the linear order parameter retains a single minimum (b), while the quadratic order parameter develops two distinct minima (d).}
    \label{fig:composite-minima}
\end{figure}

The thermal fluctuations of the order parameter are also influenced by the anisotropy parameter $\beta_-$ in the free energy density (Eq.~\eqref{eq:freeEn}), as illustrated in Fig.~\ref{fig:composite-minima}.
Panels (a) and (b) compare the free energy density profile in the $x$-$y$ plane for $\beta_- > 0$ and $\beta_- < 0$, respectively.
While the potential shape differs slightly, both cases exhibit a clear minimum at $\vec{X} = 0$.
However, the behavior changes significantly when considering $X_x^2 - X_y^2$, as shown in panels (c) and (d).
For $\beta_- > 0$ (Fig.~\ref{fig:composite-minima}(c)), $X_x^2 - X_y^2$ is minimized at $X_x^2 - X_y^2 = 0$.
In contrast, for $\beta_- < 0$ (Fig.~\ref{fig:composite-minima}(d)), the potential energy is minimized when $X_x^2 - X_y^2 \neq 0$, with a maximal value of $X_x^2 - X_y^2 \sim k_B T$.

This behavior signals the emergence of quadrupoling and the formation of a phase transition characterized by the order parameter $\langle X_1^2 - X_2^2 \rangle$ or $\langle 2X_3^2 - X_1^2 - X_2^2 \rangle$ at temperatures above the critical temperature of the vectorial order $\left<\vec{X}\right>$.
Notably, a negative $\beta_-$ does not destabilize the free energy density.
In Appendix~\ref{app:free-energy}, we demonstrate that $\beta_- < 0$ can arise from a theory with strictly positive fourth-order terms through symmetrization.

\section{Quadrupoling above the critical temperature of the parent phase}
\label{sec:quadrupoling}
In thermal equilibrium, the partition function associated with the free energy density in Eq.~\eqref{eq:freeEn} is given by
\begin{equation}
    Z = \int \mathcal{D}\vec{X} \exp{\left[ -\beta \int d\vec{r} \, f(\vec{X}, T) \right]}.
\end{equation}
We focus on temperatures above the critical temperature of the order parameter, $T > T_c$, where $\left<\vec{X}\right> = 0$.
In this regime, the fourth-order contributions to the free energy density are small.
Thus, we approximate the partition function by neglecting the isotropic fourth-order term, assuming
\begin{equation}
    Z \approx \int \mathcal{D}\vec{X} \left[1 - \beta \beta_+ \vec{X}^4 \right] \exp{\left( -\beta \int d\vec{r} \, f_g(\vec{X}, T) \right)},
\end{equation}
and consider only the zeroth-order contribution.
Here, $f_g$ represents the Gaussian part of the free energy density, $f_g = \vec{X} \left( -c \nabla^2 + r \right) \vec{X}$.

To address the fourth-order anisotropy term, we introduce auxiliary fields $\Phi_{E_g;i}$ that transform according to the basis functions of the two-dimensional irreducible representation $E_g$ of the cubic group $O_h$ ($i = 1, 2$).
This transformation is expressed as
\begin{equation}
    [\vec{X} \mathbb{M}_i \vec{X}]^2 \rightarrow \Phi_{E_g;i} \, [\vec{X} \mathbb{M}_i \vec{X}].
\end{equation}
We achieve this using the Hubbard-Stratonovich transformation, as detailed in Appendix~\ref{app:HubStr}.
This transformation allows us to rewrite the quartic interaction of the fields $X_i$ as a quadratic interaction, resulting in an overall Gaussian theory in $X_i$.
Since $T > T_c$, the values of $X_i$ are thermally fluctuating and can be integrated out, yielding an effective free energy density in terms of $\Phi_{E_g;i}$ (see Appendix~\ref{app:HubStr} for details)
\begin{multline}
    Z = \int \mathcal{D}\Phi_{E_{g;1}} \mathcal{D}\Phi_{E_{g;2}} \\
    \exp{\left[ -\beta \int d\vec{r} \, f^{\text{eff.}}(\Phi_{E_{g;1}}, \Phi_{E_{g;2}}, T) \right]},
\end{multline}
where the effective free energy density is given by
\begin{equation}
    f^{\text{eff.}} = \frac{1}{2\beta_-} \left( \Phi_{E_{g;1}}^2 + \Phi_{E_{g;2}}^2 \right) + \ln\left( \beta^{-\frac{3}{2}} \det\left( G^{-1} \right)^{-\frac{3}{2}} \right).\label{eq:feff}
\end{equation}
Here, we define the Green's functions as
\begin{align}
    G^{-1} &= G_0^{-1} + \Delta, \\
    G_0^{-1} &= c \nabla^2 + r, \\
    \Delta &= \pm i \sum_i \mathbb{M}_i \Phi_{E_{g;i}}.
\end{align}
To evaluate the Green's function term in the effective free energy density (Eq.~\eqref{eq:feff}), we use the identity $\ln(\det(G^{-1})) = \text{tr}(\ln(G^{-1}))$.
We further express $G^{-1} = G_0^{-1}(G_0 \Delta + 1)$, enabling us to use the general expansion of the natural logarithm~\cite{altland2010condensed},
\begin{equation}\label{eq:tr-ln-expansion}
    \text{tr}\left( \ln(G^{-1}) \right) = \text{const.} - \text{tr}\left[ \sum_{n=1}^{\infty} \frac{(-1)^{n+1}}{n} \left( G_0 \Delta \right)^n \right].
\end{equation}
The first term of the expansion ($n=1$) vanishes because $\text{tr}[\Delta] \sim \text{tr}[\mathbb{M}_i] = 0$.
The second-order term yields
\begin{equation}
    -\frac{1}{2} \text{tr}\left[ \left( G_0 \Delta \right)^2 \right] = -\frac{1}{2} \sum_{i, \vec{q}} \Gamma_{\vec{q}} \Phi_{E_{g;i}}^2(\vec{q}),
\end{equation}
where the vortex function is defined as
\begin{equation}
    \Gamma_{\vec{q}} = \int \frac{\mathrm{d}^3 p}{(2\pi)^3} \, G_0(\vec{p}) G_0(\vec{p} + \vec{q}).
\end{equation}
The resulting effective free energy takes the form
\begin{equation}\label{eq:free-energy}
    F\left[ \vec{\Phi}_{E_g} \right] = \sum_{\vec{q}} \left[ \frac{1}{2\beta_-} \left| \vec{\Phi}(\vec{q})_{E_g} \right|^2 + \frac{3}{4 \beta} \Gamma_{\vec{q}} \vec{\Phi}_{E_g}^2(\vec{q}) \right],
\end{equation}
where $\vec{\Phi}_{E_g} = \left( \Phi_{E_{g;1}}, \Phi_{E_{g;2}} \right)$.

\section{Self-consistent Solution of the Quadrupole Order}
\label{sec:gap-equation}
Using the effective free energy in Eq.~\eqref{eq:free-energy}, we derive a self-consistent equation for the quadrupole order parameter $\vec{\Phi}_{E_g}$. The solution to this equation determines the transition temperature of the quadrupole phase and its dependence on material parameters.

\begin{figure}[t]
    \centering
    \includegraphics[width=\linewidth]{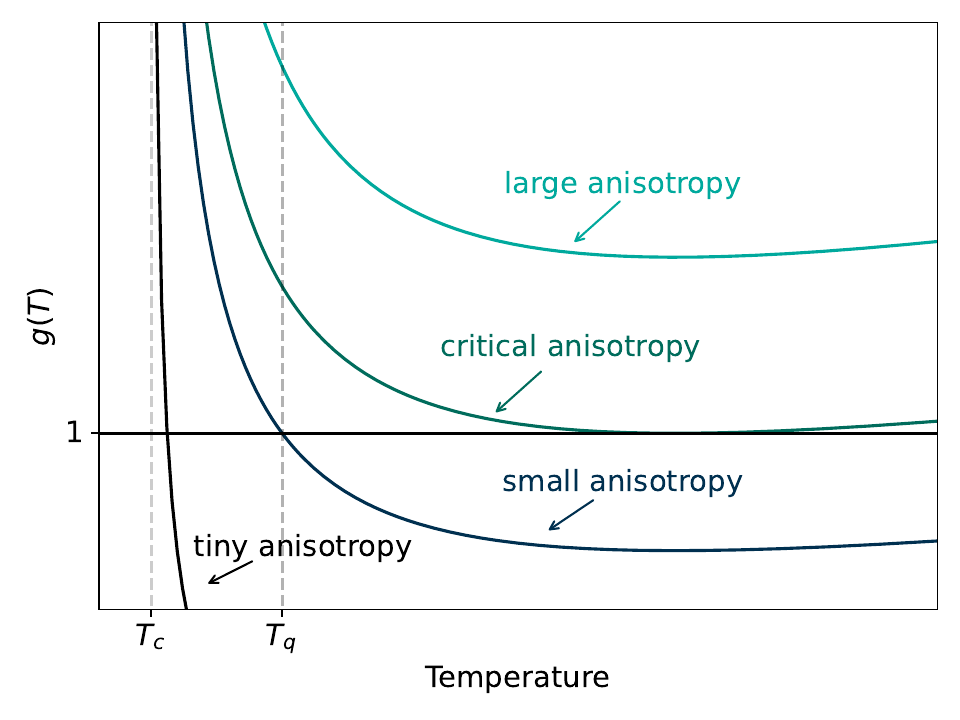}
    \caption{Gap equation for quadrupole transition. The function $g(T)$ denotes the right-hand side of the gap equation Eq.~\eqref{eq:gap-equation}. Depending on the anisotropy, the gap equation has no solutions ($\beta_- < \beta_-^*$), one solution at $T_q = 2T_c$ ($\beta_- = \beta_-^*$), or two solutions with $T_c < T_q < 2T_c$ ($\beta_- > \beta_-^*$). For a tiny anisotropy, $T_q \sim T_c$.}
    \label{fig:quadrupole-anisotropy}
\end{figure}

By applying the saddle-point approximation, we obtain the self-consistent equation for the two-component order parameter,
\begin{equation}
    \frac{\delta F\left[\vec{\Phi}_{E_g}\right]}{\delta \Phi_{E_{g;1}}} = 0\quad \text{and} \quad \frac{\delta F\left[\vec{\Phi}_{E_g}\right]}{\delta \Phi_{E_{g;2}}} = 0.
    \label{eq:saddlepoint}
\end{equation}
Since the free energy in Eq.~\eqref{eq:free-energy} depends equally on both components of the order parameter, the saddle-point approximation yields the same form for each component,
\begin{equation}
    0 = \int \frac{d^3q}{\left(2\pi\right)^3} \left[ \frac{1}{\beta_-} + \frac{3}{2\beta} \, \Gamma_q \right]\Phi_{E_{g;i}}(\vec{q}).
    \label{eq:fundev}
\end{equation}
Assuming an isotropic solution that is constant in real space, $\Phi_{E_{g;i}} = \phi_0 \delta(\vec{q})$, we find
\begin{equation}
    0 = \frac{1}{\beta_-} + \frac{3}{2\beta} \Gamma_{\vec{0}}.
    \label{eq:s-wave-gap}
\end{equation}
Here, $\vec{q} = \vec{0}$, and the calculation of the vortex function $\Gamma_{\vec{0}}$ is detailed in App.~\ref{app:vortex}. Eq.~\eqref{eq:s-wave-gap} only admits a solution if $\beta_-$ is negative, as the vortex function $\Gamma_{\vec{0}}$ is always positive. This imposes a constraint on the anisotropy of the system. Rearranging the self-consistent equation, we obtain
\begin{equation}
    1 = -\beta_- \, \frac{3}{16\pi} \frac{k_B T}{\sqrt{c^3 \alpha (T - T_c)}}.
    \label{eq:gap-equation}
\end{equation}
The solution to this equation yields the transition temperature of the quadrupole phase, $T_q$. Since the equation is proportional to $\frac{1}{\sqrt{T - T_c}}$, it exhibits a singularity at $T = T_c$, ensuring that $T_q > T_c$. The maximum value of $T_q$ is determined by the minimum of the right-hand side of Eq.~\eqref{eq:gap-equation}, which numerically corresponds to $T_q \leq 2T_c$. This follows from the condition
\begin{align}
    \frac{\partial}{\partial T} \left[ -\beta_- \, \frac{3}{16\pi} \frac{k_B T}{\sqrt{c^3 \alpha (T - T_c)}} \right] \overset{!}{=} 0.
\end{align}
Although the temperature that minimizes the gap is universally $2T_c$, the existence of a solution - i.e., an intersection between the temperature-dependent curve and the constant value of one - depends on the anisotropy of the system, encoded in $\beta_-$ in the numerator of Eq.~\eqref{eq:gap-equation}. As illustrated in Fig.~\ref{fig:quadrupole-anisotropy}, three cases can be distinguished.

For strong anisotropy, $|\beta_-| > |\beta_-^{\text{crit}}|$, Eq.~\eqref{eq:gap-equation} has no solution, and the quadrupole transition is absent. 
The critical value for the anisotropy is obtained by setting $T_q = 2T_c$ in Eq.~\eqref{eq:gap-equation}, yielding
\begin{equation}
    \beta_-^{\text{crit}} = -\frac{8\pi}{3k_B} \sqrt{\frac{c^3 \alpha}{T_c}}.
\end{equation}
Systems with this critical anisotropy have a single solution at $T_q = 2T_c$.

For weaker anisotropy, $|\beta_-| < |\beta_-^{\text{crit}}|$, the self-consistent equation has a solution $T_q < 2T_c$. 
In this case, the solution is given by
\begin{equation}
    T_q = \frac{\alpha - \sqrt{\alpha \left(\alpha - \chi^2 T_c \beta_-^2\right)}}{2\chi^2 \beta_-^2},
\end{equation}
where $\chi = \frac{3k_B}{16\pi c^{3/2}}$. 
For further decreasing anisotropy, the quadrupole transition temperature approaches the critical temperature of the dipole order, $T_q \approx T_c$.
In that case, the quadrupole phase is masked by the parent phase and no quadrupolar phase transition emerges. 
Note that we focus on the lower temperature of the two potential solutions of equation \eqref{eq:gap-equation}. 

\section{Free energy of quadrupole order}
\label{sec:quadru-free-energy}
The Landau free energy density in Eq.~\eqref{eq:feff} was truncated after the second order (compare Eq.~\eqref{eq:free-energy}). 
Higher-order terms for the free energy can be obtained from the expansion of the natural logarithm in Eq.~\ref{eq:tr-ln-expansion}, as detailed in App.~\ref{app:trace}.
\begin{figure}[t]
    \centering
    \includegraphics[width=\linewidth]{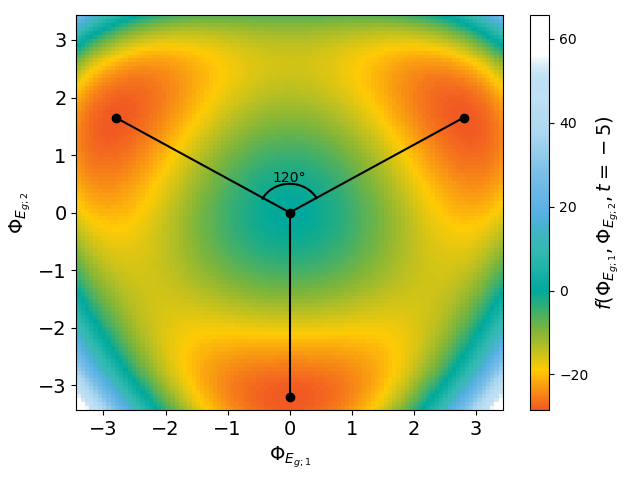}
    \caption{Minima of the quadrupole free energy density. The three distinct minima are rotationally invariant. The reduced temperature is $t=-5$, with parameters $a_0=1$, $\gamma_q = 0.4$, and $u_q = 0.6$.}
    \label{fig:minima}
\end{figure}
The free energy density up to fourth order in the quadrupole order parameter takes the form
\begin{multline}\label{eq:free-energy-phi}
    f(\Phi_{E_{g;1}},\Phi_{E_{g;2}}) = \left( \nabla \vec{\Phi}_{E_g}\right)^2 + \alpha_q(T-T_q)\vec{\Phi}_{E_g}^2 \\
    -\gamma_q \Phi_{E_{g;2}}\left(3 \Phi_{E_{g;1}}^2 -\Phi_{E_{g;2}}^2\right) + u_q \vec{\Phi}_{E_g}^4,
\end{multline}
where $\vec{\Phi}_{E_g}^4 = \Phi_{E_{g;1}}^4 + \Phi_{E_{g;1}}^2\Phi_{E_{g;2}}^2 + \Phi_{E_{g;2}}^4$.
This free energy density contains coupling terms between the different components of the order parameter, encoding asymmetries in the system.

To minimize the free energy density, we apply the mean-field approximation, setting $\nabla \vec{\Phi}_{E_g} = 0$ and introducing the reduced temperature $t = \frac{T-T_q}{T_q}$.
The mean-field free energy density follows from Eq.~\eqref{eq:free-energy-phi}
\begin{multline}\label{eq:mean-field-free-energy-quadru}
    f(\Phi_{E_{g;1}},\Phi_{E_{g;2}}) \approx a_0 t \vec{\Phi}_{E_g}^2 + u_q \vec{\Phi}_{E_g}^4 \\
    -\gamma_q \Phi_{E_{g;2}}\left(3 \Phi_{E_{g;1}}^2 -\Phi_{E_{g;2}}^2\right),
\end{multline}
where $a_0 = \alpha_q T_q$.

The order parameter of composite order is a two-component order parameter, and the extrema points are two-dimensional, with three minima found (App.~\ref{app:minima}).
Fig.~\ref{fig:minima} shows that the positions of the three minima are connected by a threefold rotation symmetry, with the depicted values for a reduced temperature $t=-5$ and parameters $a_0=1$, $\gamma_q = 0.4$, and $u_q = 0.6$.
The following discussion focuses on one minimum but applies similarly to all three.

From the minimization detailed in App.~\ref{app:minima}, we focus on the minimum with $\Phi_{E_{g;1}}=0$ and
\begin{equation}\label{eq:phi2}
    \Phi_{E_{g;2}} = \left[0, -\frac{3 \gamma_q}{8 u_q} \pm \sqrt{\left(\frac{3 \gamma_q}{8u_q}\right)^2 - \frac{a_0 t}{2 u_q}}\right].
\end{equation}
The non-trivial solution is valid only if $t < t^*=\frac{2 u_q (3\gamma_q)^2}{a_0 (8u_q)^2}$; otherwise, the order parameter would become imaginary.
This bound also defines the transition temperature for entering a non-trivial quadrupole phase.

\begin{figure}[t]
    \centering
    \includegraphics[width=\linewidth]{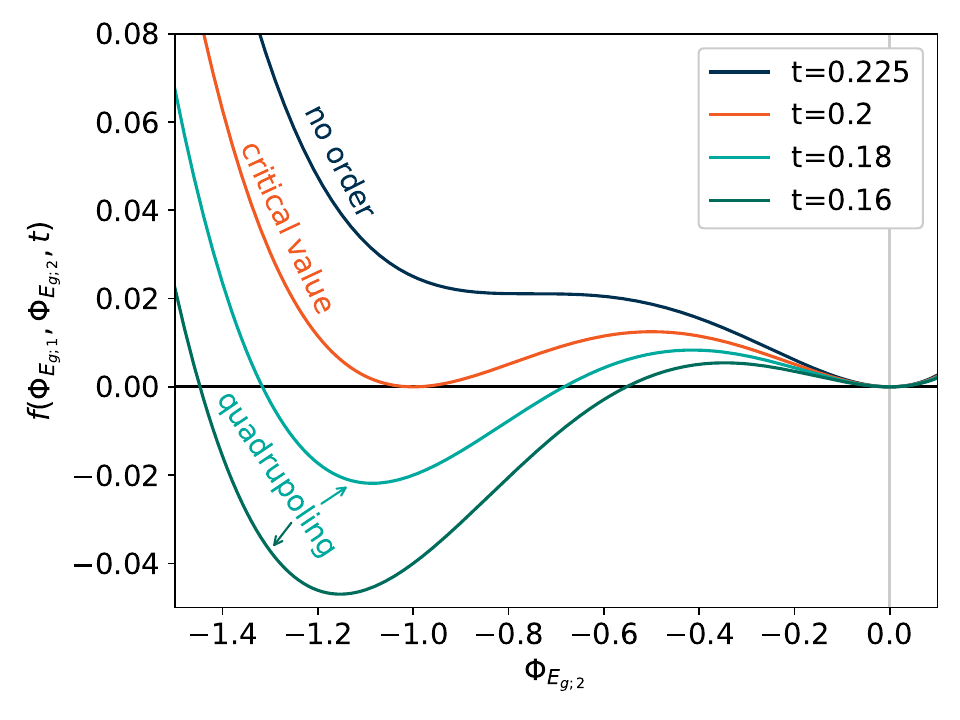}
    \caption{First-order quadrupole phase transition. The minimum of the free energy density in Eq.~\eqref{eq:free-energy-phi} continuously approaches zero with decreasing relative temperature. At the finite value of $t^*=0.2$, the quadrupole phase is entered when the minimum crosses $f(\Phi_{E_{g;1}}, \Phi_{E_{g;2}})=0$ and becomes negative. Here, $\Phi_{E_{g;1}} = 0$.}
    \label{fig:quadrupole-phasetransition}
\end{figure}

Fig. \ref{fig:quadrupole-phasetransition} shows that the free energy density for $\Phi_{E_{g;1}}=0$ always has a (local) minimum at $\Phi_{E_{g;2}}=0$.
For $t>t^*$, this minimum is the global minimum of the free energy density.
However, for $t<t^*$, a new global minimum emerges at the finite value $|\Phi_{E_{g;2}}|\neq0$.
This behavior is characteristic of a first-order phase transition, a direct consequence of the cubic term in the free energy density. Although previous theoretical work often claims to observe second-order phase transitions~\cite{ishitobi2025three,fiore2024interplay,yamamura2006structural,suzuki2018first}, definitive evidence remains lacking.

\section{Induced lattice distortion due to quadrupole order}
\label{sec:strain}
The quadrupole order parameter is even under time-reversal and inversion, enabling it to couple to mechanical strain in the material. To describe a resulting lattice distortion, we extend the free energy density in Eq.~\eqref{eq:free-energy-phi} to include the elastic free energy density, $f_{\text{elastic}}$, and the coupling of strain to the quadrupole order, $f_{\text{coupl}}$,
\begin{equation}
    \label{eq:straintotalf}
    f_\text{total} = f_\text{quad} + f_{\text{elastic}} + f_{\text{coupl}}.
\end{equation}
In cubic symmetry, the elastic free energy density up to fourth order is well-established~\cite{brassington1982cubic,brassington1983vibrational,liakos1984fourth,saunders1986thermodynamics} and summarized in Appendix~\ref{app:strain}.
Focusing on the part transforming like the irreducible representation $E_g$, the elastic free energy density in the absence of other strain components is given by
\begin{multline}
    f_{\text{elastic}} = \frac{1}{4}\left(C_{11}-C_{12}\right)\left(\eta_1^2+\eta_2^2\right) \\
    +\frac{1}{24\sqrt{3}}\left(C_{111}-3C_{112}+2C_{123}\right)\eta_2\left(\eta_2^2-3\eta_1^2\right) + \dots,
    \label{eq:felastic}
\end{multline}
where $\eta_1 = \eta_{11} - \eta_{22}$ corresponds to a shear in the $\langle 1\overline{1}0 \rangle$ direction on the $\{110\}$ plane, and $\eta_2 = \left(2\eta_{33} - \eta_{22} - \eta_{11}\right)/\sqrt{3}$ describes a tetragonal distortion in the $\langle 001 \rangle$ direction with conserved volume. Here, $C_{ij}$ and $C_{ijk}$ are the second- and third-order elastic coefficients, respectively.

The strains described by $\eta_1$ and $\eta_2$ couple linearly to the quadrupole order parameter:
\begin{equation}
    f_{\text{coupl}} = \alpha \left(\eta_1 \Phi_{E_{g;1}} + \eta_2 \Phi_{E_{g;2}}\right).
\end{equation}
In the absence of applied stress $\sigma_i$, the strain derivative of the total free energy density must vanish,
\begin{equation}
    \sigma_i = \frac{\partial f_{\text{total}}}{\partial \eta_i} \stackrel{!}{=} 0.
    \label{eq:stress}
\end{equation}
Near the quadrupole phase transition, where the induced strain is small, we focus on the quadratic terms in Eq.~\eqref{eq:felastic}. From Eq.~\eqref{eq:stress}, we obtain
\begin{equation}
    \label{eq:strainfinal}
    \eta_i = \frac{2c}{C_{11}-C_{12}} \Phi_{E_{g;i}}.
\end{equation}
Hence, shear strain and tetragonal distortion are proportional to the quadrupole orders $\Phi_{E_{g;1}}$ and $\Phi_{E_{g;2}}$, respectively. The proportionality factor depends on the phenomenological coupling strength $c$ and the second-order elastic coefficients. 

\section{Application to the double perovskite $\text{Ba}_2\text{ReMgO}_6$}
\label{sec:BaReMgO}
\begin{figure}
    \centering
    \includegraphics[width=\linewidth]{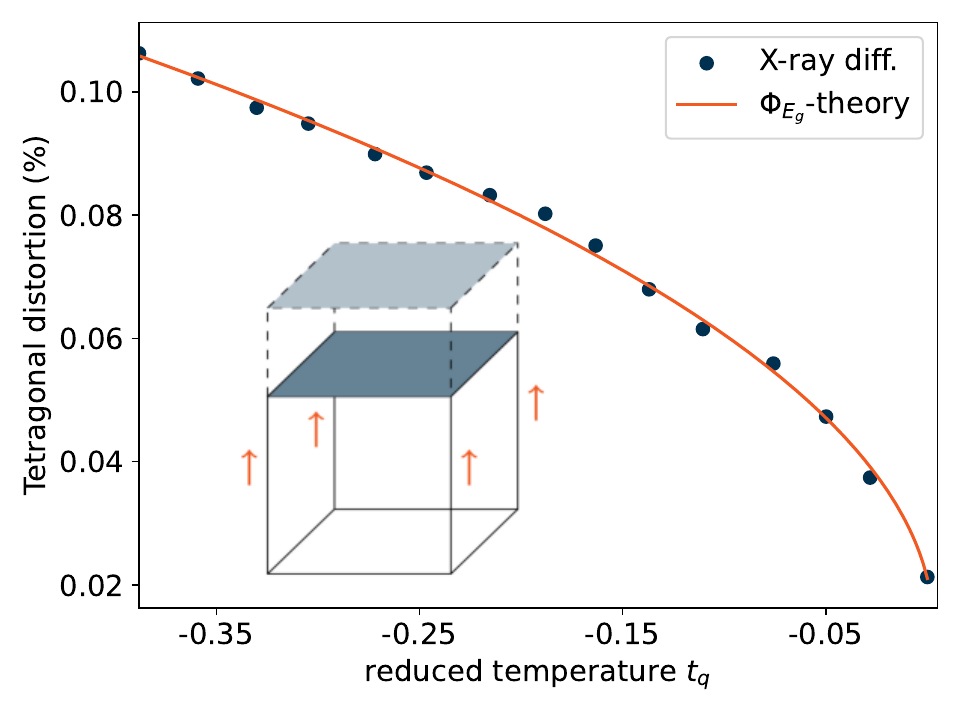}
    \caption{Tetragonal distortion of the cubic lattice symmetry of \ce{Ba2ReMgO6} induced by strain during the quadrupole phase transition. The analytical strain curve is calculated using Eq.~\eqref{eq:distortion-phi2}, and the experimental X-ray diffraction data are from Ref.~\cite{hirai2020detection}. The reduced temperature $t_q = t + t^*$ is derived from the experimental data.}
    \label{fig:strain}
\end{figure}

To illustrate our results, we discuss the coupling between strain and the quadrupolar order parameter for the double perovskite \ce{Ba2ReMgO6}.
A volume-conserving tetragonal distortion associated with quadrupole order in \ce{Ba2ReMgO6} emerges at $T_q = 33$~K~\cite{hirai2020detection}.
Synchrotron X-ray diffraction measurements reveal a continuous transition from the high-symmetry tetragonal space group $I4/mmm$ to the tetragonal space group $P4_2/mnm$, corresponding to a distortion in the $c$-direction of the \ce{ReO6} octahedron, as shown in Fig.~\ref{fig:strain}.
A second phase transition at $T_c = 18$~K introduces magnetic ordering without altering the lattice structure.

The tetragonal distortion of \ce{Ba2ReMgO6} is described by the $\eta_2$ strain component. Using the strain in Eq.~\eqref{eq:strainfinal} and the solution for $\Phi_{E_{g;2}}$ in Eq.~\eqref{eq:phi2}, we obtain
\begin{equation}\label{eq:distortion-phi2}
    \eta_2 = \frac{2c}{C_{11}-C_{12}} \left(\frac{-3 \gamma_q}{8 u_q} + \sqrt{\left(\frac{3 \gamma_q}{8u_q}\right)^2 - \frac{a_0 t}{2 u_q}}\right).
\end{equation}
The temperature dependence is governed by $a_0t$, while $u_q$ and $\gamma_q$ characterize the symmetry of the quadrupole phase. This solution for $\Phi_{E_{g;2}}$ is valid only if $\Phi_{E_{g;1}} = 0$, implying the absence of shear stress ($\eta_1 = 0$).

Since direct experimental measurements for $\gamma_q$ and $u_q$ are unavailable, we fit these parameters to the X-ray diffraction data for \ce{Ba2ReMgO6}. The results, plotted in Fig.~\ref{fig:strain}, show excellent agreement between the distortion calculated using our $\Phi_{E_g}$-theory and the experimental data, with the distortion increasing as the square root of the decreasing temperature. However, recent resonant elastic X-ray scattering measurements indicate that both $\Phi_{E_{g;2}}$ and $\Phi_{E_{g;1}}$ can coexist~\cite{Soh2024}. In our formalism, this scenario would correspond to another minimum in the free energy density, Eq.~\eqref{eq:mean-field-free-energy-quadru}, beyond the one discussed in Eq.~\eqref{eq:phi2}.

\section{Discussion}
We presented a statistical-field theory approach for a mesoscopic description of quadrupole order.
In contrast to frameworks based on competing orders, we demonstrated that quadrupole orders can emerge as composite orders in the vicinity of a dipole order parameter.
We derived the gap equation and showed that only systems with sufficient anisotropy exhibit quadrupole order.
Furthermore, we established that the transition into quadrupole order is linked to mechanical strain, leading to tetragonal distortions or shear in cubic systems, consistent with experimental observations~\cite{hirai2020detection,Soh2024}.

We view the mesoscopic approach presented here as complementary to previously reported microscopic theories.
Its applicability relies on the assumption of a multicomponent order parameter, which naturally emerges in crystal structures of sufficiently high symmetry.
While we used cubic symmetry as the primary example in this manuscript, the same procedure can be applied to other symmetries.
For example, in tetragonal systems, an in-plane magnetization could be connected to spin nematic order, similar to the discussion for the two-component superconducting state~\cite{fernandes2019intertwined}.

Additionally, as mentioned in the introduction, extending this approach beyond magnetic systems could provide insights into the softening behavior reported for several ferroelectrics above the critical temperature.
For example, in PbSc$_{0.5}$Ta$_{0.5}$O$_3$, a precursor regime with softening of the shear elastic constant was identified, extending over $\approx 100$~K above the ferroelectric transition~\cite{Aktas2013}.

Our derivation of quadrupolar order assumes that the anisotropy term in the free energy density dominates over the isotropic term, which we disregarded.
Furthermore, we provided a phenomenological discussion of the quadrupolar order using mean-field theory.
It would be interesting to extend this formalism by accounting for order parameter fluctuations, both of the primary and quadrupole orders, for example, through numerical Monte Carlo simulations or the renormalization group formalism~\cite{hasenbusch2023cubic,hasenbusch2024phi}.
In line with the conventional theory of critical phenomena, we expect that fluctuations would lower the critical temperature for the quadrupole phase.

Additionally, investigating the dynamic properties of the quadrupole order would be of interest.
For example, driving the shear mode would couple to the quadrupolar order, potentially giving rise to anomalous heating behavior above the critical temperature.
The heat production in this scenario is expected to be $\Delta Q \sim \int \mathrm{d}t\, \left<\eta_i(t) \Phi_{E_g;i}\right>$~\cite{Tietjen2025,caprini2024ultrafast}.

\medskip 

\section*{Acknowledgements}
We acknowledge support from the Swedish Research Council (VR starting Grant No. 2022-03350), the Olle Engkvist Foundation (Grant No. 229-0443), the Royal Physiographic Society in Lund (Horisont), the Knut and Alice Wallenberg Foundation (Grant No. 2023.0087), as well as the department of physics and the areas of advance Nano and Material Science at Chalmers University of Technology. 
\bibliography{references}

\appendix

\section{Free Energy to Partition Function} \label{app:free-energy}
We begin with the Landau free energy density of $\vec{X}$ for cubic symmetry~\cite{geilhufe2018gtpack,barabanov2022landau,devonshire1954theory}, where $\vec{X}$ represents either magnetization or polarization,
\begin{align}
    f(\vec{X},T) = &f_0(T) + c \left(\nabla \vec{X}\right)^2 + \alpha(T-T_c)\vec{X}^2 \nonumber \\
    &+ \beta_1 \left(X_x^4 + X_y^4 + X_z^4\right) \nonumber \\
    &+ \beta_2 \left(X_x^2X_y^2 + X_y^2X_z^2 + X_z^2X_x^2\right). \label{app:f_en}
\end{align}
The anisotropy depends on the material-specific coefficients $\beta_1$ and $\beta_2$, the critical temperature is given by $T_c$. 

The goal is to re-express the free energy density \eqref{app:f_en} in terms of bilinears transforming as the 2D irreducible representation $E_g$ of the cubic group. To do so, we introduce 
\begin{align}
    \phi_{E_{g;1}} &= \frac{1}{\sqrt{2}}(X_x^2 - X_y^2), \\
    \phi_{E_{g;2}} &= \frac{1}{\sqrt{6}}(X_x^2 + X_y^2 - 2X_z^2).
\end{align}
For notational convenience, we define the matrices $\mathbb{M}_1$ and $\mathbb{M}_2$:
\begin{equation}
    \mathbb{M}_1 = \frac{1}{\sqrt{2}} \begin{bmatrix}
    1 & 0 & 0 \\
    0 & -1 & 0 \\
    0 & 0 & 0
    \end{bmatrix}, \quad
    \mathbb{M}_2 = \frac{1}{\sqrt{6}} \begin{bmatrix}
    1 & 0 & 0 \\
    0 & 1 & 0 \\
    0 & 0 & -2
    \end{bmatrix},
\end{equation}
which allow us to express the bilinears as $\phi_{E_{g;1}} = \vec{X} \mathbb{M}_1 \vec{X}$ and $\phi_{E_{g;2}} = \vec{X} \mathbb{M}_2 \vec{X}$.

We also define the following parameters:
\begin{align}
    \beta_+ &= \frac{\beta_1 + \beta_2}{3}, \\
    \beta_- &= 2\beta_1 - \beta_2, \\
    r &= \alpha(T - T_c).
\end{align}
Note that the definition of $\beta_-$ implies the possibility that $\beta_-<0$, although $\beta_1, \beta_2 > 0$.

Using these notations, the reformulated free energy density takes the form:
\begin{multline}
    f(\vec{X}, T) = \vec{X} \left( -c \nabla^2 + r \right) \vec{X} + \beta_+ \vec{X}^4 \\
    + \frac{\beta_-}{2} \left([\vec{X} \mathbb{M}_1 \vec{X}]^2 + [\vec{X} \mathbb{M}_2 \vec{X}]^2\right) .
\end{multline}

\section{Hubbard-Stratonovich transformation to quadrupole field\label{app:HubStr}}
To remove the fourth-order terms in the free energy density, we introduce quadrupolar order fields $\Phi_{E_g;i}$ using the Hubbard-Stratonovich transformation to the partition function,
\begin{align}
    \label{eq:partfunc-hubbard}
    Z &= \int \mathcal{D}\vec{X} \exp{\left( -\beta\int d\vec{r} \, f(\vec{X}, T) \right)} \\
    &= \int \mathcal{D}\vec{X} \int \mathcal{D}\Phi_{E_{g;1}} \int \mathcal{D}\Phi_{E_{g;2}} \nonumber \\
    &\quad \exp\left[-\beta \int d\vec{r} \, \vec{X}\left(-c\nabla^2 + r \pm i\mathbb{M}_1\Phi_{E_{g;1}} \pm i\mathbb{M}_2\Phi_{E_{g;2}} \right) \vec{X} \right. \nonumber \\
    &\quad \left. + \beta_+ \vec{X}^4 + \frac{1}{2\beta_-} (\Phi_{E_{g;1}}^2 + \Phi_{E_{g;2}}^2) \right].
\end{align}
Here, the Green's functions are defined as
\begin{align}
    G^{-1} &= G_0^{-1} + \Delta, \\
    G_0^{-1} &= c\nabla^2 + r, \\
    \Delta &= \pm i \sum_i \mathbb{M}_i \Phi_{E_{g;i}}.
\end{align}

Above the critical point, $\vec{X}$ fluctuates, and we assume the isotropic fourth-order term $\beta_+ \vec{X}^4$ to be negligible. 
We can now use the Gaussian integral with source term,
\begin{equation*}
    \int_{-\infty}^{\infty} \exp{\left(-\frac{1}{2a} x^2 \pm i bx\right)} dx = \sqrt{\frac{2 \pi}{a}} \exp{\left(-\frac{1}{2} ab^2\right)}.
\end{equation*}
For the Gaussian integral to be applicable, $a > 0$ is required.
After performing the Gaussian integral in Eq.~\eqref{eq:partfunc-hubbard}, we obtain the partition function,
\begin{multline}
    \label{eq:app:partition-final}
    Z = \int \mathcal{D}\Phi_{E_{g;1}} \int \mathcal{D}\Phi_{E_{g;2}} \\ \exp{\left[ -\beta \int d\vec{r} \, \frac{1}{2\beta_-} (\Phi_{E_{g;1}}^2 + \Phi_{E_{g;2}}^2) \right]} \\
    \times \exp\left[ \ln\left(\beta^{-\frac{3}{2}} \det\left(G^{-1}\right)^{-\frac{3}{2}}\right) \right].
\end{multline}

\section{Expansion of the natural logarithm of the Green's function}\label{app:trace}
The natural logarithm is expanded into a generalized power series,
\begin{equation}
    \text{tr}\left(\ln(G^{-1})\right) \sim - \text{tr}\left[\sum_{n=1}^{\infty} \frac{(-1)^{n+1}}{n} \left(G_{0} \Delta\right)^{n}\right].
\end{equation}
We evaluate the trace explicitly by summing over all $k$-space. Note that $G_0$ is diagonal, i.e., $\langle\vec{k}|G_0|\vec{q}\rangle = G_0(\vec{k})\delta_{\vec{k},\vec{q}}$.
For $n=1$, the trace vanishes because $\text{tr}\left[\Delta\right] \sim \text{tr}\left[\mathbb{M}_i \right] = 0$. For $n=2$, we obtain,
\begin{align}
    \text{tr}&\left[\left(G_0\Delta\right)^2\right] = \sum_{\vec{k}} \langle \vec{k}|G_0\Delta G_0\Delta |\vec{k}\rangle \nonumber \\
    &= \sum_{i,\vec{k},\vec{l}} G_0(\vec{k}) \Phi_{E_{g;i}}(\vec{k}-\vec{l}) G_0(\vec{l})\Phi_{E_{g;i}}(\vec{l}-\vec{k}) \nonumber \\
    &\quad \left(\text{with } \vec{q} = \vec{k}-\vec{l}, \vec{p} = \vec{l} \rightarrow \vec{k} = \vec{q}+\vec{p}\right) \nonumber \\
    &= \sum_{i,\vec{q},\vec{p}} G_0(\vec{q}+\vec{p})G_0(\vec{p}) \Phi_{E_{g;i}}(\vec{q})\Phi_{E_{g;i}}(-\vec{q}) \nonumber \\
    &= \sum_{i, \vec{q}} \Gamma_{\vec{q}} \Phi_{E_{g;i}}^2(\vec{q}),
\end{align}
where $\Phi_{E_{g;i}} \in \mathbb{R}$ and $\Gamma_{\vec{q}} = \sum_{\vec{p}} G_0(\vec{q}+\vec{p})G_0(\vec{p})$.

For higher-order terms, we proceed similarly. For $n=3$, we get:
\begin{align}
    \text{tr}&\left[\frac{\left(G_0\Delta\right)^3}{3}\right] = \frac{1}{3}\sum_{\vec{k}} \langle \vec{k}|G_0\Delta G_0\Delta G_0\Delta |\vec{k}\rangle \nonumber \\
    &= \frac{1}{3\sqrt{6}} \sum_{\vec{k}, \vec{l}, \vec{m}} G_0(\vec{k})G_0(\vec{m}+\vec{k})G_0(\vec{l}+\vec{k}) \Phi_{E_{g;2}}(\vec{l}) \nonumber \\
    &\quad \times \left[ 3\,\Phi_{E_{g;1}}(\vec{m})\Phi_{E_{g;1}}(\vec{m}-\vec{l}) - \Phi_{E_{g;2}}(\vec{m})\Phi_{E_{g;2}}(\vec{m}-\vec{l})\right],
\end{align}
where we used $\text{tr}\left[ \mathbb{M}_2 \mathbb{M}_1^2\right] = \frac{1}{\sqrt{6}}$ and $\text{tr}\left[ \mathbb{M}_2^3\right] = -\frac{1}{\sqrt{6}}$, while other third-order combinations of matrices vanish.

For $n=4$, we use $\text{tr}\left[ \mathbb{M}_1^2 \mathbb{M}_2^2\right] = \frac{1}{6}$, $\text{tr}\left[ \mathbb{M}_2^4 \right] = \frac{1}{2}$, and $\text{tr}\left[ \mathbb{M}_1^4 \right] = \frac{1}{2}$.
Applying the same steps, we obtain:
\begin{align}
    \text{tr}&\left[\frac{-\left(G_0\Delta\right)^4}{4}\right] = \frac{-1}{8}\sum_{\substack{i\neq j, \\\vec{k}, \vec{l}, \\ \vec{m}, \vec{p}}} G_0(\vec{k})G_0(\vec{k}-\vec{p})\Phi_{E_{g;i}}(\vec{k})\Phi_{E_{g;i}}(\vec{k}-\vec{p}-\vec{l}) \nonumber \\
    &\quad \times \left[ G_0(\vec{l})G_0(\vec{l}+\vec{m}) \Phi_{E_{g;i}}(\vec{m})\Phi_{E_{g;i}}(\vec{m}+ \vec{l}- \vec{k}) \right. \nonumber \\
    &\quad \left. + G_0(\vec{l})G_0(\vec{l}+\vec{m}) \Phi_{E_{g;j}}(\vec{m})\Phi_{E_{g;j}}(\vec{m}+ \vec{l}- \vec{k})\right].
\end{align}

\section{Vortex function} \label{app:vortex}
From the expansion of the natural logarithm in the free energy density (Eq.~\eqref{eq:feff}), we obtain in second order,
\begin{align}
    -\frac{1}{2}\text{tr}\left[\left(G_0\Delta\right)^2\right] &= -\frac{1}{2}\sum_{i,\vec{q},\vec{p}} G_0(\vec{q})G_0(\vec{p}+\vec{q}) \Phi_{E_{g;i}}(\vec{q})\Phi_{E_{g;i}}(-\vec{q}) \nonumber \\
    &= -\frac{1}{2}\sum_{i, \vec{q}} \Gamma_{\vec{q}} \Phi_{E_{g;i}}^2(\vec{q}),
    \label{eq:tr-ln-n1}
\end{align}
where the vortex function is defined as,
\begin{multline}
    \label{eq:app:vortex-function}
    \Gamma_{\vec{q}} = \int \frac{\mathrm{d}^3p}{(2\pi)^3} G_0(\vec{p})G_0(\vec{p}+\vec{q}) \\
    = \int \frac{\mathrm{d}^3p}{(2\pi)^3} \frac{1}{\left( \frac{1}{2} c \vec{p}^2 + \frac{1}{2}r\right)\left( \frac{1}{2} c (\vec{p} +\vec{q})^2 + \frac{1}{2}r\right)}.
\end{multline}
To calculate the vortex function, we use the Feynman trick~\cite{Peskin2018} to rewrite the convolution integral over the Green's functions,
\begin{align}
    \Gamma_{\vec{q}} &= \int_0^1 \mathrm{d}x \int \frac{\mathrm{d}^3p}{(2\pi)^3} \frac{1}{\left[(1-x)\left(c \vec{p}^2 + r\right) + x\left(c (\vec{p}+\vec{q})^2 + r\right)\right]^2} \nonumber \\
    &= \int_0^1 \mathrm{d}x \int \frac{\mathrm{d}^3p}{(2\pi)^3} \frac{4}{c^2\left[\vec{p}^2 + 2x\vec{p}\cdot\vec{q} + x\vec{q}^2 + \frac{r}{c}\right]^2} \nonumber \\
    &= \int_0^1 \mathrm{d}x \int \frac{\mathrm{d}^3l}{(2\pi)^3} \frac{1}{c^2\left[\vec{l}^2 + x(1-x)\vec{q}^2 + \frac{r}{c}\right]^2},
\end{align}
where the last equality is obtained by setting $\vec{p} = \vec{l} - x\vec{q}$.

The integrand is spherically symmetric in $\vec{l}$, so transforming to spherical coordinates yields an integral with only radial dependence,
\begin{equation}
    \Gamma_{\vec{q}} = \int_0^1 \mathrm{d}x \int \frac{\mathrm{d}^3l}{2\pi^2} \frac{l^2}{c^2\left[\vec{l}^2 + x(1-x)\vec{q}^2 + \frac{r}{c}\right]^2}.
\end{equation}

We use the Gamma-function identity,
\begin{equation}
    \int_0^{\infty} \mathrm{d}l\, \frac{l^2}{\left(l^2 + a\right)^2} = \frac{\pi}{4} \frac{1}{a^{1/2}},
\end{equation}
where $a = x(1-x)\vec{q}^2 + \frac{r}{c}$. For $d=3$ and $\alpha=2$, the relevant Gamma-function values are $\Gamma(2) = 1$, $\Gamma(1/2) = \sqrt{\pi}$, and $\Gamma(3/2) = \sqrt{\pi}/2$. Consequently, the vortex function becomes,
\begin{equation}
    \Gamma_{\vec{q}} = \frac{1}{8\pi c^{3/2}} \int_0^1 \mathrm{d}x \frac{1}{\sqrt{x(1-x)\vec{q}^2 + \frac{r}{c}}}.
\end{equation}

\section{Minima of Quadrupole Free Energy} \label{app:minima}
To compute the minima of the quadrupole free energy density in Eq.~\eqref{eq:mean-field-free-energy-quadru}, we solve the self-consistent equations by setting the derivatives with respect to each order parameter component to zero,
\begin{equation}
    \frac{\delta F\left[\vec{\Phi}_{E_g}\right]}{\delta \Phi_{E_{g;i}}} = 0, \quad i = 1, 2.
\end{equation}
This yields three minima in total, as shown in Fig.~\ref{fig:minima} of the main text.

The analytical solutions for the first component are
\begin{equation}
    \Phi_{E_{g;1}}^* = \left[0, \pm \frac{\sqrt{-a_0 t + \Phi_{E_{g;2}}(3 \gamma_q - u_q \Phi_{E_{g;2}})}}{\sqrt{2 u_q}} \right],
\end{equation}
where $t = \frac{T-T_q}{T_q}$ is the reduced temperature.

For $\Phi_{E_{g;1}} = 0$, we calculate the exact analytical form of the minimum by solving,
\begin{equation}
    2 a_0 t \Phi_{E_{g;2}} + 3 \gamma_q \Phi_{E_{g;2}}^2 + 4 u_q \Phi_{E_{g;2}}^3 = 0,
\end{equation}
which yields
\begin{equation}
    \Phi_{E_{g;2}} = \left[0, \frac{-3 \gamma_q}{8 u_q} \pm \sqrt{\left(\frac{3 \gamma_q}{8 u_q}\right)^2 - \frac{a_0 t}{2 u_q}}\right].
\end{equation}
The second solution is valid only if the reduced temperature is below its critical value, $t < t^* = \frac{2 u_q (3 \gamma_q)^2}{a_0 (8 u_q)^2}$. 

\section{Fourth-order invariants for the Landau free-energy expansion of strain in cubic symmetry} \label{app:strain}
Up to fourth order, the elastic free energy density in cubic symmetry is expressed as\cite{brassington1982cubic}
\begin{align*}
    \varphi = &\frac{1}{6} \left( C_{11} + 2C_{12} \right) \left( \eta_0^0\right)^2 + \frac{1}{4} \left( C_{11} - C_{12} \right) \left( \eta_1^2 + \eta_2^2\right) 
    \\ &+ \frac{1}{2} C_{44}\left( \eta_3^2 + \eta_4^2 + \eta_5^2 \right) + \frac{1}{54} \left( C_{111} + 6C_{112} + 2C_{123} \right) \left( \eta_0^0\right)^3 \\
    &+ \frac{1}{12} \left( C_{111} - C_{123} \right) \eta_0^0 \left( \eta_1^2 + \eta_2^2\right) \\
    &+ \frac{1}{24 \sqrt{3}} \left( C_{111} -3 C_{112} + 2 C_{123} \right) \eta_1 \left( \eta_1^2 - 3 \eta_2^2\right) \\
    &+ \frac{1}{6} \left( C_{144} + 2C_{155} \right) \eta_0^0 \left( \eta_3^2 + \eta_4^2 + \eta_5^2\right) \\
    &+\frac{1}{4 \sqrt{3}} \left( C_{144} - C_{155} \right) \eta_1\left( 2\eta_5^2 - \eta_4^2- \eta_3^2\right)^2 \\
    &+ \frac{1}{4} \left( C_{144} -C_{155} \right) \eta_2\left( \eta_3^2 - \eta_4^2\right)^2 + C_{456}\eta_3 \eta_4 \eta_5.
    \end{align*}
Here, the strain tensor components transforming as the irreducible representations of the cubic group are given by 
\begin{align*}
    \eta_0^0 &= \eta_{11} + \eta_{22} +\eta_{33}\, , &A_{1g}\\
    \eta_1 &= \left( 2\eta_{33} -\eta_{22} -\eta_{11} \right)/ \sqrt{3}\, , &E_g\\ \eta_1' &= \left( 2\eta_{11} -\eta_{33} -\eta_{22} \right)/ \sqrt{3}\, , & E_g\\
    \eta_1'' &=  \left( 2\eta_{22} -\eta_{33} -\eta_{11} \right)/ \sqrt{3}\, , &E_g\\
    \eta_2 &= \eta_{11}-\eta_{22}\, , \quad \eta_2' = \eta_{22}- \eta_{33}\, , \quad \eta_2'' = \eta_{33}- \eta_{11} \, , &E_g \\
    \eta_3 &= \eta_{23}\,,\quad \eta_4 = \eta_{13}\,,\quad \eta_5=\eta_{12} &T_{2g}
\end{align*}

\end{document}